\documentclass[utf8]{FrontiersinHarvard}
\usepackage{url,hyperref,lineno,microtype,subcaption}
\usepackage{datetime}
\usepackage{fmtcount}
\usepackage{etoolbox}
\usepackage{fcprefix}
\usepackage[onehalfspacing]{setspace}
\usepackage{graphicx}
\usepackage{bm}                      
\usepackage{mathptmx}                
\usepackage{xcolor}                  
\usepackage{amssymb,amsmath}
\usepackage{multirow}
\usepackage{ulem,xpatch}
\nolinenumbers

\def\keyFont{\fontsize{8}{11}\helveticabold}
\def\firstAuthorLast{G. F. Burgio {et~al.}} 
\def\Authors{First Author\,$^{1,*}$, Co-Author\,$^{2}$ and Co-Author\,$^{1,2}$}
\def\Address{$^{1}$Laboratory X, Institute X, Department X, Organization X, City X , State XX (only USA, Canada and Australia), Country X} 
\def\corrAuthor{Corresponding Author}

\def\corrEmail{email@uni.edu}

\def\bea{\begin{eqnarray}} \def\eea{\end{eqnarray}}
\def\beq{\begin{equation}} \def\eeq{\end{equation}}
\def\bal#1\eal{\begin{align}#1\end{align}}
\def\bse#1\ese{\begin{subequations}#1\end{subequations}}

\def\text{\mathrm}

\def\ms{M_\odot}
\def\mmax{M_\text{max}}
\def\esnm{E_\text{SNM}}

\def\esym{E_\text{sym}}

\def\fm3{\;\text{fm}^{-3}}

\begin{document}

\onecolumn
\firstpage{1}

\title[Symmetry energy and skin thickness]{The nuclear symmetry energy and the neutron skin thickness in nuclei} 

\author[\firstAuthorLast ]{\Authors} 
\address{} 
\correspondance{} 

\def\Authors{G. F. Burgio\,$^{1,*}$, H. C. Das\,$^{1}$ and I. Vida\~na\,$^{1}$}
\def\Address{$^{1}$INFN Sezione di Catania, and Department of Physics and Astronomy, Università di Catania, Via S. Sofia 64, Catania, Italy}
\def\corrAuthor{G. F. Burgio}

\def\corrEmail{fiorella.burgio@ct.infn.it}
\extraAuth{}
\maketitle

\begin{abstract}

We investigate possible correlations between the stiffness of the symmetry energy at saturation density, the so-called $L$ parameter, and the neutron skin thickness of ${^{48}}$Ca and ${^{208}}$Pb, for which the recent measurements from the CREX and PREX I+II experiments at the Thomas Jefferson Laboratory became available.
We choose an ensemble of nucleonic equations of state (EoS) derived  within microscopic (BHF, Variational, AFDMC) and phenomenological (Skyrme, RMF, DD-RMF) approaches. They are all compatible with the laboratory nuclear collisions data and with current observations of neutron stars (NS) mass and the tidal polarizability of a 1.4 $M_{\odot}$ NS, as deduced from the GW170817 event.
We find some degree of correlation between the $L$ parameter and the neutron skin thickness whereas a much weaker correlation does exist with the tidal polarizability and the symmetry energy at saturation density. However, some EoS which are able to explain the CREX experimental data, are not compatible with the PREX I+II data, and viceversa.
We confirm the results previously obtained with a different set of EoS models, and find a possible tension between the experimental data and the current understanding of the nuclear EoS.
\tiny{
{ \keyFont{\section{Keywords:} neutron star, equation of state, many-body methods of nuclear matter, neutron skin thickness, CREX, PREX I+II.}}}
\end{abstract}

\section{Introduction}
\label{sec:intro}
The nuclear symmetry energy plays a major role on the structure of neutron-rich finite nuclei as well as on the bulk properties of neutron stars \citep{EPJA_2014,2016PrPNP}. In the past decades, several laboratory experiments \citep{Russotto_2023} 
have been performed in order to investigate the symmetry energy in finite nuclei, e.g. measurements of the nuclear masses \citep{moller2012}, the nuclear dipole polarizability \citep{rocavin2015}, the giant and pygmy dipole resonance energies \citep{pygmy1,pygmy2}, isobaric analog states \citep{danlee2014}, and the neutron skin thickness \citep{Adhikari_2021, Adhikari_2022}. Several neutron stars (NS) properties are sensitive to the symmetry energy, e.g. its internal composition, the crust-core transition density and therefore the crust thickness, and the presence of fast direct URCA neutrino processes which regulate NS cooling \citep{Yakovlev_2004, Burgio_2021}.

The symmetry energy is directly related to the more general and comprehensive task of the study of the equation of state (EoS), which plays a major role in  nuclear structure studies, analysis of the heavy-ion collision dynamics, and the physics of compact objects \citep{oertel,2018Chap6}. The central density likely reached in NS interiors is about one order of magnitude larger than the nuclear saturation density, and this poses serious problems in theoretical astrophysics, because a  correct theory of nuclear interactions for highly dense matter should be derived from the quantum chromodynamics (QCD). The well-known sign problem of lattice QCD still bars access to the high-density EoS, and therefore, models extracted from the nuclear many-body theory are required in order to build the EoS. Predictions have to be tested both in terrestrial laboratories, and in astrophysical observations. The most promising NS observables are the mass and radius; as far as the masses are regarding, the ones of several NSs are known with good precision \citep{mass,demorest2010,heavy2,fonseca2016, cromartie, Romani_2022}, while the information on their radii \citep{ozel16,gui2013} has been improved thanks to the combined observations of NICER \citep{nicer3,millernicer} and Advanced LIGO and Virgo collaborations, with the detection of gravitational waves emitted during the GW170817 NS merger event \citep{merger,mergerl,mergerx}. This event has provided us with important new information on the NS mass and radii by means of the measurement of the tidal deformability \citep{hartle,flan}, thus deducing upper and lower limits on it \citep{mergerl,radice}. Further constraints on mass and radius have been recently reported by NICER for PSR J1231-1411, having mass $M= 1.04^{+0.05}_{-0.03}$ \citep{Salmi:2024bss}.

In this paper, we concentrate on the study of the neutron skin thickness $\delta R$ in neutron-rich nuclei, such as $\rm ^{208}Pb$ and $\rm ^{48}Ca$, which has long been recognized as being strongly dependent on the slope of the symmetry parameter $L$. Novel data on $\rm ^{208}Pb$ (PREX-I and PREX-II) \citep{Adhikari_2021} and $\rm ^{48}Ca$ (CREX) \citep{Adhikari_2022} with direct measurements consisting of parity-violating and elastic electron scattering technique \citep{skin3}, recently became available from the Thomas Jefferson Laboratory. Correlations between the neutron skin thickness, symmetry parameters, and NS observables like the radius of 1.4$M_\odot$ and tidal polarizability have been widely analyzed, see, e.g., \citep{Lattimer_2023} and references therein. 

In our previous paper \citep{universe6080119}, we studied those kind of correlations by choosing a set of EoS based on microscopic methods and phenomenological approaches, and discussing their behaviour with respect to the PREX-I experimental data, which were the available ones at that time. Now, we would like to elaborate more on that analysis, taking into account the recent PREX-II \citep{Adhikari_2021} and CREX data \citep{Adhikari_2022}. Moreover, we now choose a set of equations of state EoS which are compatible with the NS data on the highest observed mass $M > 2.14^{+0.10}_{-0.09} M_{\odot}$ and tidal polarizability of a 1.4 $M_{\odot}$, $\Lambda_{1.4}=190^{+390}_{-120}$. This way, we should be able to improve our study on the possible correlations among observational quantities and 
properties of nuclear matter close to saturation.

The paper is organized as follows. In Sect.\ref{sec:2} we illustrate some basic  properties of the EoS adopted in this work, along with the criteria selection for the choice of the optimal EoS. The laboratory and observational constraints on the nuclear EoS are presented in Sect.\ \ref{sec:3}. The neutron skin thickness is discussed in Sect.\ \ref{sec:4}, and conclusions are drawn in Sect.\ \ref{sec:5}.
\section{The nuclear equation of state}
\label{sec:2}
The composition of high density nuclear matter currently represents one of the most intriguing issues in theoretical physics, and several and diverse predictions have been proposed thus far \citep{Burgio_2021}. The description of the extreme density conditions can include different scenarios, e.g., a purely nucleonic one characterized by a large neutron-proton asymmetry, or hyperonic matter or a hadron-quark phase transition. All those issues suffer of drawbacks that the current experimental data, either heavy-ion collisions in terrestrial laboratories or NS observations, cannot solve. In this work, we assume that nucleons are the most relevant degrees of freedom. 

Theoretical approaches to determine the nuclear EoS are usually classified in microscopic and phenomenological ones. The interested reader is referred to recent reviews  \citep{2018Chap6,Burgio_2021}; in this paper, we skip details and summarize the main features of the adopted methods. For the microscopic approaches, we adopt several EoS derived in the Brueckner--Hartree--Fock (BHF) theory \citep{baldo1999}, which is based on the use of realistic two- and three-body forces (TBF), derived from meson-exchange theory \citep{bonn1,nij1} 
and describe correctly the nucleonic phase shifts and the properties of the deuteron. For the TBF we use the phenomenological Urbana model \citep{uix1,uix2,uix3}, and a microscopic TBF \citep{glmm,uix3,zuo1,tbfnij,li08}. We adopt as nucleon-nucleon potentials the Bonn B \citep{bonn1,bonn2}, the Nijmegen 93 \citep{nij1,nij2}, and the Argonne $V_{18}$ \citep{v18}, which are supplemented by microscopic TBF and labeled in the following as BOB, N93 and V18. The Urbana model has been used with the Argonne $V_{18}$ potential and is labeled as UIX. The explicit inclusion of the quark-gluon degrees of freedom in the construction of a potential model has been performed in Ref.\ \citep{2014PhRvL.113x2501B,2015PhRvC..92f5802F}, in which case two different EoS versions labeled respectively as FSS2CC and FSS2GC in Table~\ref{t:sat} have been obtained. Besides BHF EoS, in this paper we exploit the often-used results of the relativistic Dirac-BHF method (DBHF) \citep{dbhf3}, which employs the Bonn~A potential, the variational APR EoS \citep{apr1998} based on the $V_{18}$ potential, and the so-called CBF-EI model, obtained within the correlated basis function approach \citep{Benhar_2017}, using a realistic nuclear Hamiltonian with the Argonne V6' \citep{Wiringa_2002}  and the Urbana IX nuclear potentials as TBF. For completeness, we also include in our set a parametrization of the Auxiliary Field Diffusion Monte Carlo (AFDMC) calculation \citep{afdmc}.  

The philosophy of the phenomenological approaches is quite different from the one characterizing microscopic methods. In fact, they are based on effective interactions that are built to describe finite nuclei in their ground state, and therefore, predictions at high isospin asymmetries and density have to be taken with care \citep{stone}. Among the most used ones, we mention Skyrme interactions \citep{skyrmea} and relativistic mean-field (RMF) models \citep{rmfbb}. In this work, we use a set of modern Skyrme EoS, which are listed in  Table~\ref{t:sat}; in particular we mention the SLy0-SLy10 \citep{sly} and SLy230a \citep{sly230a1,sly230a2} of the Lyon group, and the BSk20, BSk25 and BSk26 of the Brussels group \citep{Potekhin_2013, Goriely2013}, the latter ones being unified EoS constructed on the basis of the energy-density functional theory. A complementary approach is given by RMF models, which are based on effective Lagrangian densities. The interaction between baryons is described in terms of meson exchanges, which are regulated by coupling constants of nucleons with mesons, and are usually fixed by fitting the bulk properties of nuclear matter as well as masses and radii of finite nuclei. In this work, we consider two types of RMF models: models with density-dependent coupling constants labeled DD-RMF \citep{ddme,tw99, Xia_2022} and RMF models with fixed coupling strength \citep{SINPA_N_B, BigApple_A, BigApple_B, BSR, FSUGarnet, FSUGZ03, G2_star, IUFSU, G3, IOPB, NITR}.

The main properties of the chosen EoS at saturation density are listed in Table~\ref{t:sat}. We notice that, whereas the saturation properties of the phenomenological models are within the empirical range, some microscopic EoS are marginally compatible with it. The reason is that the parameters of the phenomenological models are fitted on the saturation properties, while they are a prediction in the case of microscopic calculations, and those depend both on the many-body approach and the choice of the employed forces.  For instance, the V18 EoS predicts a slightly too low $E_0$ and $K_0$, which is mainly due to the inclusion of a particular TBF \citep{tbfnij}. We stress that a complete ab-initio theory of TBF is not available yet.

For completeness, we remind that the above mentioned methods are suited for describing the homogeneous component of the nuclear matter EoS, and that at densities 
$\rho < \rho_t \approx 0.08\,\text{fm}^{-3}$, clusters have to be included for the description of the NS crust. For that, we use the well-known Negele-Vautherin EoS \citep{NV} in the density range 
($0.001\,\text{fm}^{-3} < \rho < 0.08\,\text{fm}^{-3}$), and the ones by Baym-Pethick-Sutherland \citep{bps} and Feynman-Metropolis-Teller \citep{fey} for densities $\rho < 0.001\,\text{fm}^{-3}$ typical of the outer crust.
\subsection{Criteria for the selection of the EoS}
\label{sec:2.1}
The most important criterium for selecting the EoS is to check its behaviour with respect to the saturation properties of nuclear matter. In fact, around saturation density $\rho_0$ and isospin asymmetry $\delta \equiv (N-Z)/(N+Z)=0$ [being $Z(N)$ the number of protons (neutrons)], the energy per particle of asymmetric nuclear matter $E(\rho,\delta)$ can be expanded as a function of density and isospin asymmetry, and the coefficients are given by a set of few isoscalar ($E_0, K_0$) and isovector ($S_0, L, K_{sym}$) parameters, which can be constrained by nuclear experiments. The expansion reads
\bea
 E(\rho,\delta) &=& \esnm(\rho) + \esym(\rho) \delta^2 \:,
\label{e:ea}
\\
 \esnm(\rho) &=& E_0 + \frac{K_0}{2} x^2 \:,
\\
 \esym(\rho) &=& S_0 + L x + \frac{K_\text{sym}}{2} x^2 \:,
\label{e:ebulk}
\eea
where $x \equiv (\rho-\rho_0)/3\rho_0$, $E_0$ is the energy per particle of symmetric nuclear matter (SNM) at $\rho_0$, $K_0$ the incompressibility and
$S_0 \equiv E_\text{sym}(\rho_0)$
is the symmetry energy coefficient at saturation, defined as
\bea
 K_0 &\equiv& 9\rho_0^2  \frac{d^2\esnm}{d\rho^2}(\rho_0) \:,
\\
 S_0 &\equiv& \frac{1}{2} \frac{\partial^2E}{\partial\delta^2}(\rho_0,0)
\eea
The density dependence of the symmetry energy around saturation is characterized by the parameters $L$ and $K_\text{sym}$, which are expressed as
\bea
 L &\equiv& 3 \rho_0 \frac{d\esym}{d\rho}(\rho_0) \:,
\\
 K_\text{sym} &\equiv& 9 \rho_0^2 \frac{d^2\esym}{d\rho^2}(\rho_0) \:.
\eea
In Table~\ref{t:sat}, we list the saturation properties of the various considered EoSs, and compare them with available experimental data. Measurements of nuclear masses \citep{audi03} and density distributions \citep{vries87} yield the saturation point $E_0=-16\pm 1$ MeV and $\rho_0=0.14-0.17$ fm$^{-3}$, whereas the value of $K_0$ can be extracted from the analysis of isoscalar giant monopole resonances in heavy nuclei, reporting $K_0=240\pm 10$ MeV \citep{colo04} or $K=248 \pm 8$ MeV \citep{piekarewicz04}, in agreement with the low value of $K_0$ found in heavy ion collision experiments\citep{fuchs01}. 
We also notice that, whereas $S_0$ is more or less well established ($\approx 30$ MeV), the values of $L$ ($30\,\text{MeV} < L < 87\,\text{MeV}$), is still quite uncertain \citep{Reed21,Essick21,Lattimer_2023}. Also $K_{sym}$ ($-400\,\text{MeV} < K_\text{sym} < 100\,\text{MeV}$) is poorly constrained \citep{tews2017,2017Zhang}.

Besides the laboratory data, we also exploit astrophysical observation of NS. A very important constraint to be fulfilled by the different EoS is the value of the maximum NS mass, which has to be compatible with the observational data \citep{demorest2010,heavy2,fonseca2016}, in particular, the recent lower limit $\mmax>2.14\pm0.1 \ M_\odot$ \citep{cromartie}. The GW detection by Advanced LIGO and Advanced Virgo \citep{merger, mergerl, mergerx} of the GW170817 event put strong constraints on the so-called tidal polarizability $\Lambda$ \citep{hinder2008,hinder2009,hinder2010}, which is strongly influenced by the EoS. The GW170817 analysis for a $1.4\,\ms$ NS \citep{merger} gave an upper limit of $\Lambda<800$, which was later improved to $\Lambda=190^{+390}_{-120}$ \citep{mergerl}.

From Table~\ref{t:sat}, we notice that most of the adopted EoSs in this work are compatible with the nuclear empirical values, the NS maximum mass, and the tidal deformability of a 1.4 $M_\odot$ NS. We, therefore, consider this set of EoSs for the analysis of the neutron skin thickness, which is discussed in the following Sect. \ref{sec:4}. 

\begin{table}[!p]
\renewcommand{\arraystretch}{1.15}
\begin{center}
\caption{
Saturation properties predicted by the considered EoSs.
Experimental nuclear parameters and observational data are listed for comparison. The references in the lower part of the table are labeled as
[a] \citep{margue2018a}, 
[b] \citep{shlomo}, 
[c]  \citep{pieka}, 
[d] \citep{2018Chap6}, 
[e]  \citep{Burgio_2021}, 
[f]  \citep{mergerl}, and 
[g] \citep{cromartie}. 
See text for details.}
\label{t:sat}
\scalebox{0.8}{
\begin{tabular}{lcccccccr}
\hline\hline
 Model class & EoS    & $\rho_0[\fm3]$ & $-E_0$[MeV] & $K_0$[MeV] & $S_0$[MeV] & $L$[MeV] & $\Lambda_{1.4}$ & $\rm M_{max}[M_\odot]$\\
\hline
 Microscopic & BOB  & 0.170 & 15.40 & 238 & 33.70 & 70.00 & 570 & 2.50 \\
 & V18     & 0.178  & 13.90 & 207 & 32.30 & 67.00 & 440 & 2.36 \\
 & N93     & 0.185  & 16.10 & 229 & 36.50 & 77.00 & 473 & 2.25\\
 & UIX     & 0.171  & 14.90 & 171 & 33.50 & 61.00 & 309 & 1.96 \\
 & FSS2CC  & 0.157  & 16.30 & 219 & 31.80 & 52.00 & 295 & 1.94  \\
 & FSS2GC  & 0.170  & 15.60 & 185 & 31.00 & 51.00 & 262 & 2.08 \\
 & DBHF    & 0.181  & 16.20 & 218 & 34.40 & 69.00 & 681 & 2.31 \\
 & APR     & 0.159  & 15.90 & 233 & 33.40 & 51.00 & 250 & 2.19 \\
 & CBF-EI  & 0.160  & 10.90 & 240 & 30.00 & 68.00 & 501 & 2.47 \\
 & AFDMC   & 0.160  & 16.00 & 239 & 31.30 & 60.00 & 256 & 2.20\\
 \hline
Skyrme & Rs & 0.158  & 15.05 &  248 &  30.83 &  86.41 & 910 & 2.27\\
& SGI  & 0.155 & 15.89 & 265 & 28.35 & 63.85 & 714 & 2.31 \\
& SLy0 &    0.160   &    16.01   &     226  &     31.40    &    45.37 & 315 & 2.06\\
& SLy1 & 0.161    &   15.98    &   233  &     32.59   &     48.88 & 314 & 2.06\\
& SLy2 & 0.161    &    15.92     &   235  &      32.39  &     48.84 & 318 & 2.06\\
& SLy3 &   0.161    &    15.96   &     233  &      32.12     &   45.56 & 295 & 2.05\\
& SLy4 &     0.160   &     15.97   &    232  &      31.85   &     45.38 & 309 & 2.06 \\
& SLy5 &   0.161  &     15.98  &      232    &    32.70  &     50.34 & 328 & 2.07\\
& SLy6 &   0.159  &      15.92  &     230 &      31.21 &       45.21 & 334 & 2.09 \\
& SLy7 &  0.159     &   15.90   &     233 &      32.41   &     48.11 & 337 & 2.09 \\
& SLy8 &    0.161 &       15.96  &     233  &     32.51    &   45.36 & 316 & 2.06\\
& SLy9 &    0.151    &   15.79   &    229 &      32.12 &      55.37 & 513 & 2.23 \\
& SLy10 &    0.156    &    15.90   &     232  &      32.19   &     39.24 & 262 & 2.04\\
& SLy230a &    0.160 &      15.98   &     230 &      31.88  &      43.99 & 340 & 2.16 \\
& SkI4 &    0.160 &       16.15   &    239  &      29.38  &    59.34& 581 & 2.29 \\
& SkMP &   0.157  &     15.57   &     230  &     29.70 &      69.70 & 666 & 2.19\\
& SkO & 0.161    &    15.78  &     228 &     32.19  &      79.92 & 656 & 2.10\\
& SkO' & 0.160  &     15.73   &    222    &   32.10 &      69.68 & 465 & 2.00\\
& SkT4 & 0.159  &      15.95  &     235 &       35.23   &     93.48 & 919 & 2.23 \\
& SkT5 & 0.164  &      15.99  &     201 &       37.60   &     100.3 & 807 & 2.08\\
& BSk20 & 0.160 & 16.04 & 241 & 30.00 & 37.40 &  328 & 2.18\\
& BSk25 & 0.158 & 15.99 & 236 & 29.00 & 36.90 & 545 & 2.22\\
& BSk26 & 0.159 & 16.03 & 240 & 30.00 & 37.50 & 332 & 2.18 \\
\hline 
RMF & SINPA & 0.151 & 16.00 & 204 &  31.24 &  54.01 &  586 & 2.00\\
& SINPB & 0.150 & 16.04 & 206 &  33.92 &  71.47 &  623 & 1.99 \\
& GL97 & 0.152 & 15.56 & 226 & 32.10 & 88.55 & 600 & 2.00 \\
& BigApple & 0.155 & 16.34  & 226 &  31.33 &  39.88 &  796 & 2.62\\
& BSR8 & 0.148 & 16.04 & 233 & 31.19 & 60.60 & 792 & 2.05 \\
& BSR9 & 0.148 & 16.07 & 234 & 31.68 & 64.07 & 793 & 2.04\\
& FSUGarnet & 0.153 & 16.23 & 229 &  30.89 & 53.85 & 740 & 2.12\\
& FSUGZ03 & 0.148 & 16.07 & 234 &  31.61 & 64.17 & 794 & 2.04 \\
& G2* & 0.153 & 15.95 & 213 & 30.29 & 69.43 & 692 & 2.05 \\
& IUFSU & 0.160 & 16.70 & 241 & 31.88 & 49.57 & 602 & 2.00\\
& G3 & 0.148 & 16.02 & 244 & 30.20 & 45.34 & 461 & 2.00 \\
& IOPB-I & 0.149 & 16.10 & 222 & 33.30 & 63.54 & 681 & 2.15 \\
& NITR & 0.155 & 16.32 & 224 & 31.51 & 43.46 & 683 & 2.36 \\
\hline
DD-RMF & DD & 0.148 & 16.50 & 239 & 32.58 & 58.73 & 750 & 2.43 \\
& DD2 & 0.148 & 16.02 & 240 & 32.03  & 58.00 & 753 & 2.44 \\
& DD-ME1 & 0.152 &  16.23 &  245 & 34.13 & 58.38 & 704 & 2.46 \\
& DD-ME2 & 0.152 &  16.14 &  251 & 33.39 & 54.25 & 766 & 2.50 \\
& TW-99 & 0.152 & 16.10 & 239 &  33.18 & 58.40 & 446 & 2.10 \\
\hline
& Exp.   & $\sim 0.14-0.17$ & $\sim 15-17$ & $220-260$ & $28.5-34.9$ & $30-87$ & $70-580$ &  $>2.14^{+0.10}_{-0.09}$\\
& Ref.  & [a]
        & [a]
        & [b]\, , \ [c]
        & [d]\, , \ [e]
        & [d]\, , \ [e]
        & [f]
        & [g] \\
\hline\hline
\end{tabular}}
\end{center}
\end{table}

\section{Constraints on the nuclear EoS}
\label{sec:3}
\begin{figure}[t]
\hspace{2mm}
\centerline{\includegraphics[scale=0.7]{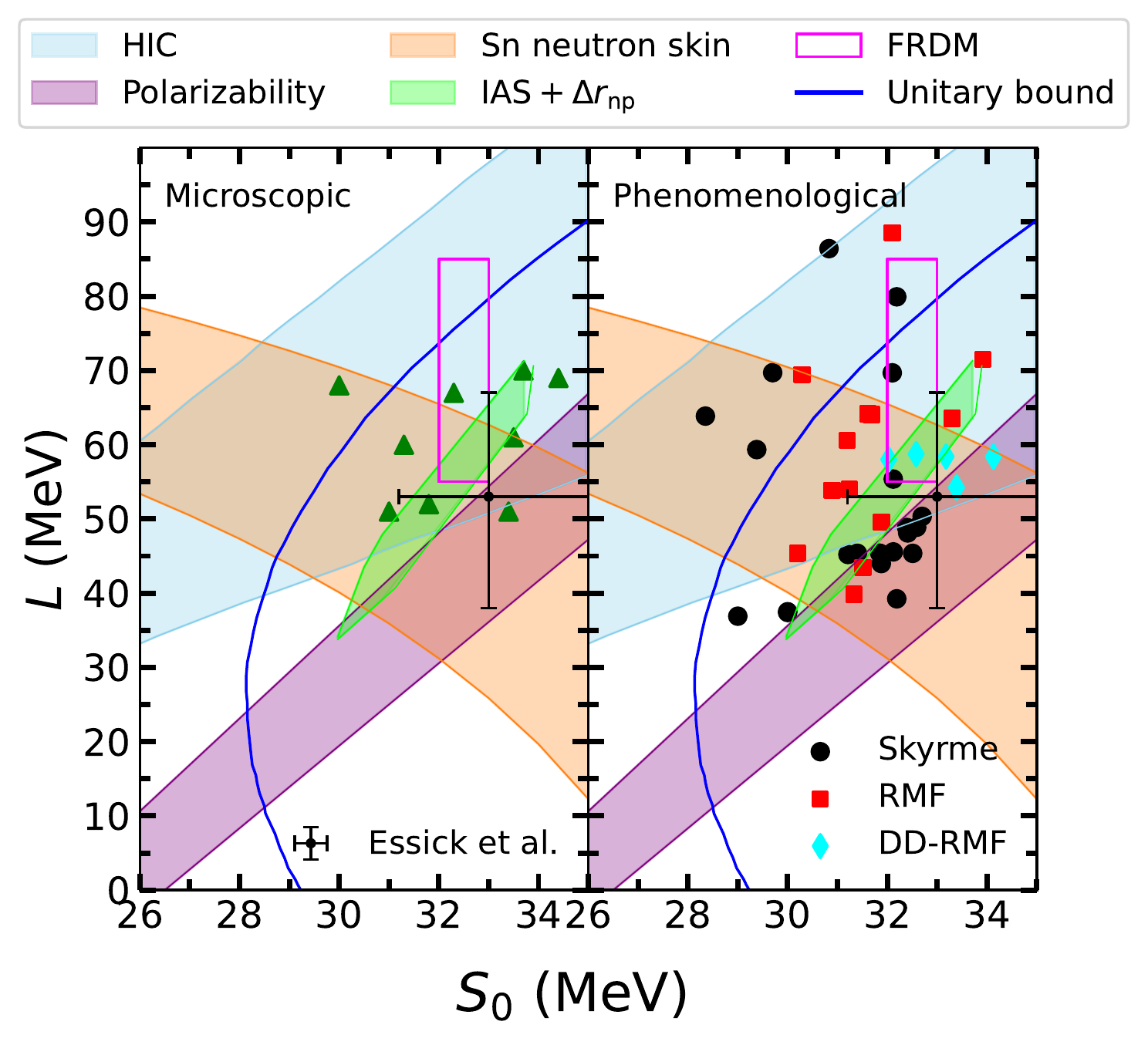}}
\vspace{-7mm}
\caption{
Symmetry energy slope $L$ vs. the symmetry energy at saturation $S_0$. 
See text for details on the experimental data.}
\label{f:fig1}
\end{figure}

An important check for the EoS is the behaviour of the symmetry energy slope $L$ vs. $S_0$, and this is plotted in Fig.~\ref{f:fig1}. The full green triangles represent the microscopic calculations (left panel), whereas the phenomenological ones are shown in the right panel. The experimental constraints indicate those derived from the study of isospin diffusion in heavy ion collisions (HIC, blue band)\citep{2009PhRvL.102l2701T} ; the electric dipole polarizability (violet band) \citep{rocavin2015}; the neutron skin thickness in Sn isotopes (orange band) \citep{chen2010}; the finite-range droplet mass model calculations (FRDM, magenta rectangle) \citep{moller2012}; the isobaric analog state (IAS) phenomenology combined with the $^{208}$Pb neutron-skin thickness (green band) \citep{danlee2014}; the recent analysis of the PREX-II experiment (black cross) \citep{Essick21}. The blue solid curve is the unitary gas bound \citep{tews2017}: only values of $(S_0,L)$ to the right of the curve are allowed. We see that all considered constraints are not simultaneously fulfilled in any area of the parameter space, probably because of the strong model dependencies in the extraction of the constraints from the raw data. Hence, at the moment, no theoretical models can be ruled out a priori, except those which are predicting values of the symmetry energy parameters outside the considered range.

\begin{figure}[h]
\hspace{2mm}
\centerline{\includegraphics[scale=0.6]{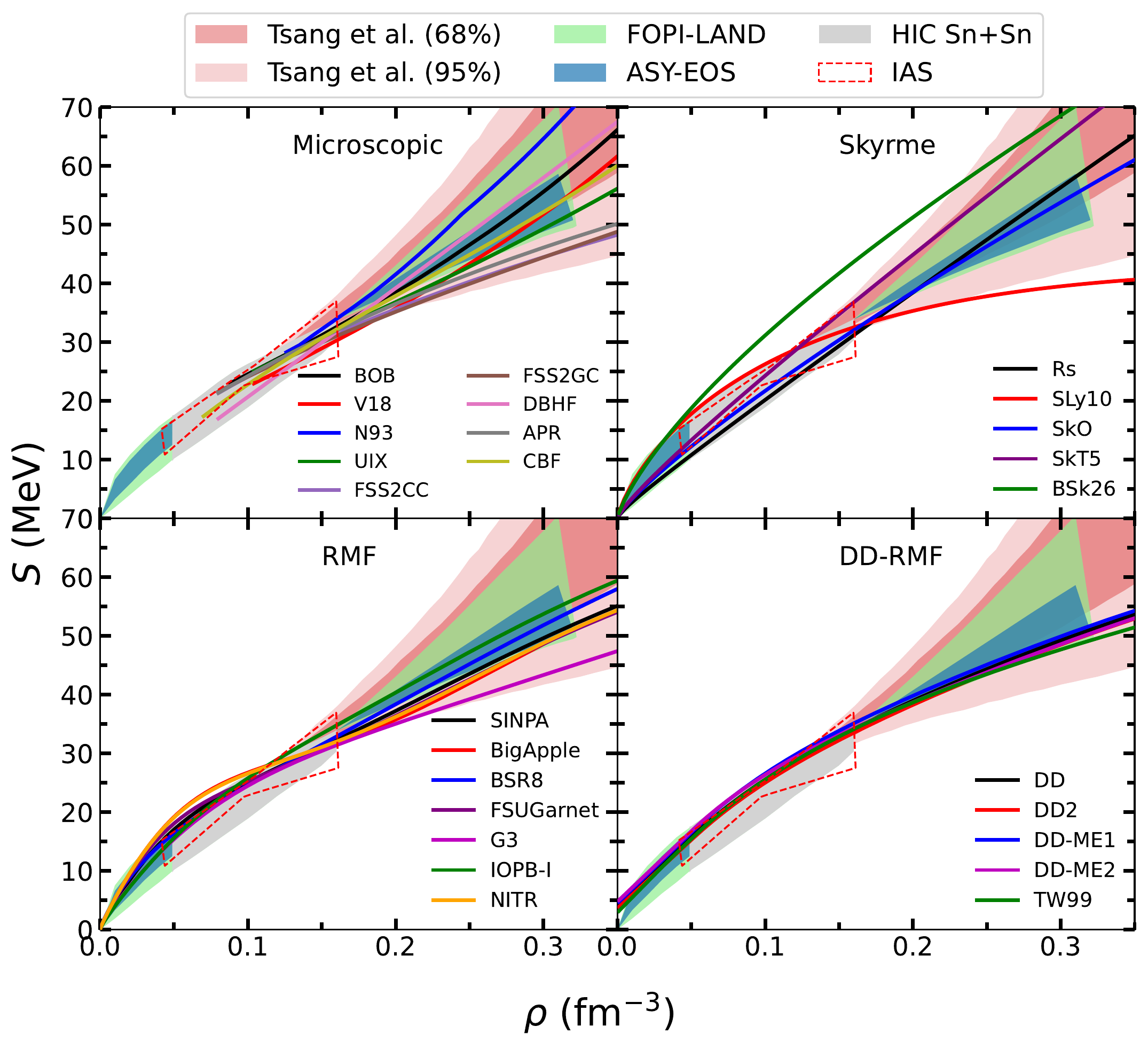}}
\vspace{-7mm}
\caption{
Symmetry energy vs. the nucleon density $\rho$. See text for details.}
\label{f:fig2}
\end{figure}

A further important check regards the high-density behaviour of the nuclear symmetry energy, as illustrated in Ref.\ \citep{Russotto_2023}. In the last few years several heavy-ion collisions experiments at relativistic energies have been performed in order to constrain the high-density symmetry energy. Fig.~\ref{f:fig2} displays the ASY-EOS data \citep{2016PhRvC..94c4608R} (blue band) and the FOPI-LAND ones \citep{2011PhLB..697..471R} (light green band) as a function of the density, HIC (Sn+Sn) diffuseness measurements \citep{2009PhRvL.102l2701T} (grey band), whereas the red dashed contour labeled by IAS shows the results of Ref. \citep{danlee2014}. For completeness, we also display the results of a Bayesian analysis \citep{Tsang_2024} which determines the boundaries at 68\% (dark pink) and 95\% confidence intervals (light pink) of the posterior distributions using an initial sample size of 3M of EoS.  The experimental data are plotted up to $\rho = 2\rho_0$, and they all show a monotonically increasing behaviour with increasing density. In the four panels the symmetry energy is plotted vs. the nucleonic density for the microscopic models (upper left), for some of the Skyrme models (upper right), RMF models (lower left), and DD-RMF models (lower right) listed in Table ~\ref{t:sat}. Except a couple of cases, i.e. BSk26 and SLy10, all EoS agree with the experimental data and the Bayesian analysis, thus confirming the need of more accurate experiments in order to disentangle the various theoretical approaches.

\section{The neutron skin thickness}
\label{sec:4}
The strong correlation between the neutron skin thickness and the slope parameter $L$ at normal nuclear saturation density was shown first by Brown and Typel \citep{skin1,skin2}, and confirmed later by other authors \citep{stei05,cen09,horo01,skin3,fur02}. A measurement of the thickness allows to establish an empirical calibration point for the pressure of neutron star matter at subnuclear densities, and coupled with a NS radius measurement could determine the pressure at supranuclear densities. In fact, the neutron skin thickness and the NS size originate both from the pressure of neutron-rich matter, hence are sensitive to the same EoS. Therefore, the Typel-Brown correlation would be helpful in establishing the pressure–density relationship over a wide range of densities inside neutron stars.

The neutron skin thickness can be defined as the difference between the neutron ($R_n$) and proton ($R_p$) root-mean-square radii: $\delta R = \sqrt{\langle r_n^2 \rangle} - \sqrt{\langle r_p^2 \rangle}$. Since the microscopic approaches discussed before are not suited for the description of finite nuclei, we prefer to use a different approach based on Ref. \citep{2009Isaac}, in which an estimation of the neutron skin thickness of $^{208}$Pb and $^{132}$Sn was made following the suggestion of Ref. \citep{stei05}. In this case $\delta R$ is calculated to lowest order in the diffuseness corrections as $\delta R \sim \sqrt{\frac{3}{5}}t$, being $t$ the thickness of semi-infinite asymmetric nuclear matter
\begin{equation}
t=\frac{\delta_c}{\rho_0(\delta_c)(1-\delta_c^2)}\frac{E_s}{4\pi r_0^2}
\frac{\int_0^{\rho_0(\delta_c)}\rho^{1/2}[S_0/E_{sym}(\rho)-1][E_{SNM}(\rho)-E_0]^{-1/2} d\rho}
{\int_0^{\rho_0(\delta_c)}\rho^{1/2}[E_{SNM}(\rho)-E_0]^{1/2}d\rho} \ .
\label{eq:t}
\end{equation}

\begin{figure}[t]
\centerline{\includegraphics[scale=0.7]{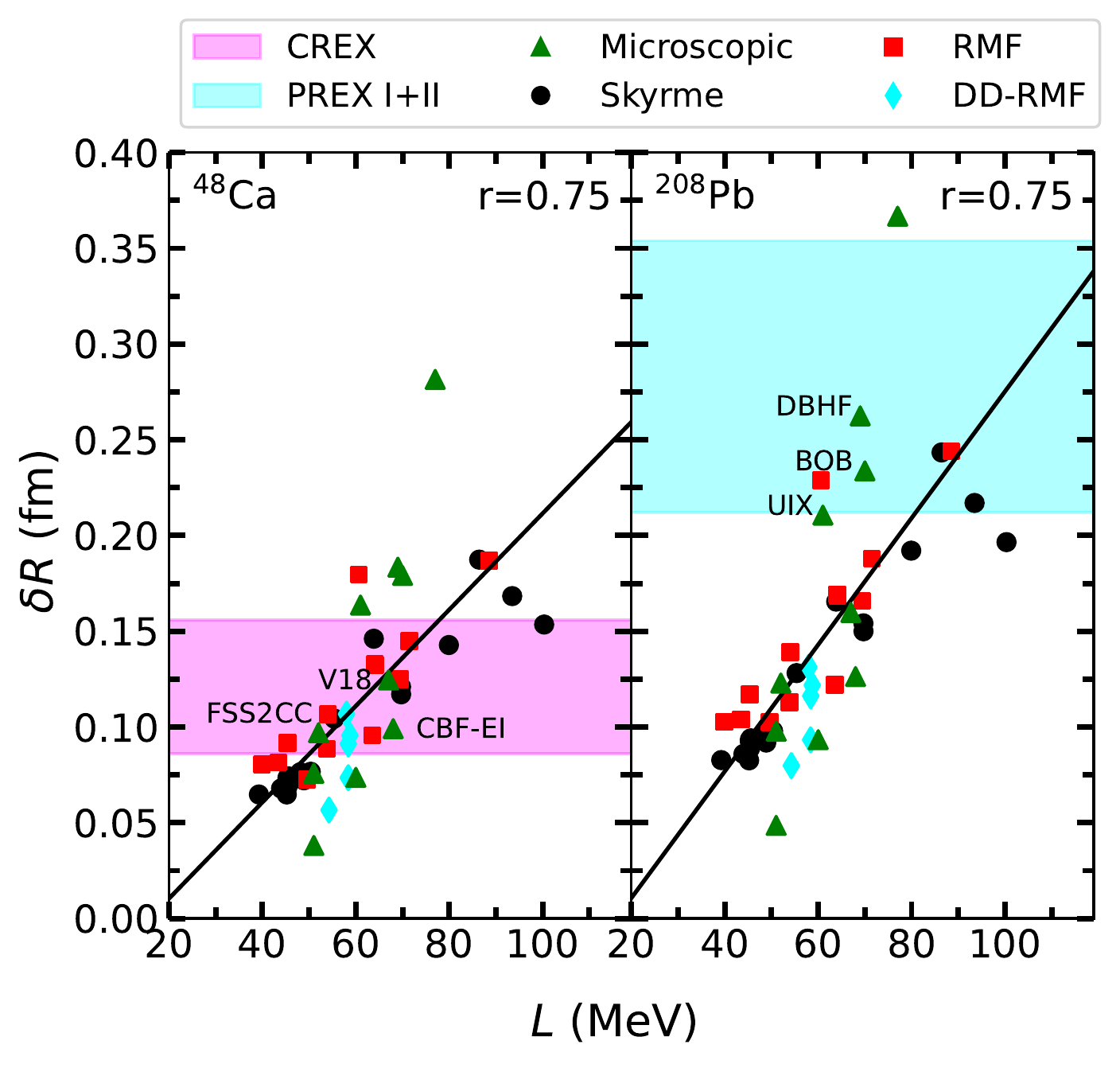}}
\vspace{4mm}
\caption{The neutron skin thickness for $\rm^{48}Ca$ (left panel) and $\rm^{208}Pb$ (right panel) is displayed as a function of $L$ for the different EoS present in Table~\ref{t:sat}.
The bands show the experimental constraints discussed in ref. \citep{Adhikari_2021, Adhikari_2022}. The solid lines indicate a linear fit of the EoS data. The values of the corresponding correlation factors $r$ are also given.}
\label{f:fig3}
\end{figure}

In that expression, $E_s$ is the surface energy taken from the semi-empirical mass formula equal to $17.23$ MeV, $r_0$ is obtained from the normalization condition $(4\pi r_0^3/3)(0.16)=1$, and $\delta_c$ is the isospin asymmetry in the center of the nucleus taken as $\delta_c=\delta/2$ according to Thomas-Fermi calculations. For consistency, we use this same method also for calculating the thickness in the phenomenological approaches. In Fig.~\ref{f:fig3}, we show the results of our calculations and compare them with experimental bands regarding CREX (left panel, magenta) and PREX I+II (right panel, cyan). Those experiments yield for PREX I+II a neutron skin thickness $\delta \rm R(^{208}Pb)=0.283 \pm 0.071 \, fm$, whereas the measurement of the neutron skin of $\rm^{48}Ca$ with the same technique gives smaller values, i.e. $\delta \rm R(^{48}Ca) =0.121 \pm 0.035 \, fm$, thus pushing $L$ towards larger or smaller values respectively. Data are shown as a function of the parameter $L$.  The linear increase of $\delta$R with $L$ is not surprising because the neutron skin thickness  in heavy nuclei is determined by the pressure difference between neutrons and protons, and this is proportional to the parameter $L$, that is, $P(\rho_0,\delta)\approx L \rho_0 \delta^2/3$.  We notice that the theoretical predictions show some correlation between $\delta R$ and $L$, as indicated by the linear fits (solid line) and by the value of the correlation coefficient, $r=0.75$, in both cases. This is slightly smaller than the previous result shown in ref. \citep{universe6080119}, where the chosen EoS set was not filtered with respect to the NS observational data. 
It has to be noticed that, in both cases, the microscopic calculations alone seem to lie on a curve with a slope different than the one of the phenomenological calculations. However, due to the small number of points available for the microscopic approaches, no firm conclusion can be drawn.
We also notice that the same EoS is unable to reproduce both the CREX and PREX I+II data; for instance, among the microscopic approaches, the CREX data set is well reproduced by the FSS2CC, V18, and CBF-EI models, whereas DBHF, BOB, and UIX fall in the range of PREX I+II. Those EoS differ not only by the many-body technique adopted, but also by the nucleon-nucleon interaction. A similar behavior can also be found for the phenomenological models such as Rs, SkT4, BSR8, and GL97 respectively. This might indicate a possible tension between the experimental data and the current understanding of the EoS. Though, it has to be stressed that the present calculations of the neutron skin thickness are based on the concept of semi-infinite nuclear matter (see Eq.\ (\ref{eq:t})), which could be inappropriate. However, a similar result has already been found in other works, using suitable methods for finite nuclei \citep{Lattimer_2023}.  Further laboratory experiments on medium size nuclei, or a re-analysis of the current data could help to clarify this point.

\begin{figure}[!htbp]
\centerline{\includegraphics[scale=0.6]{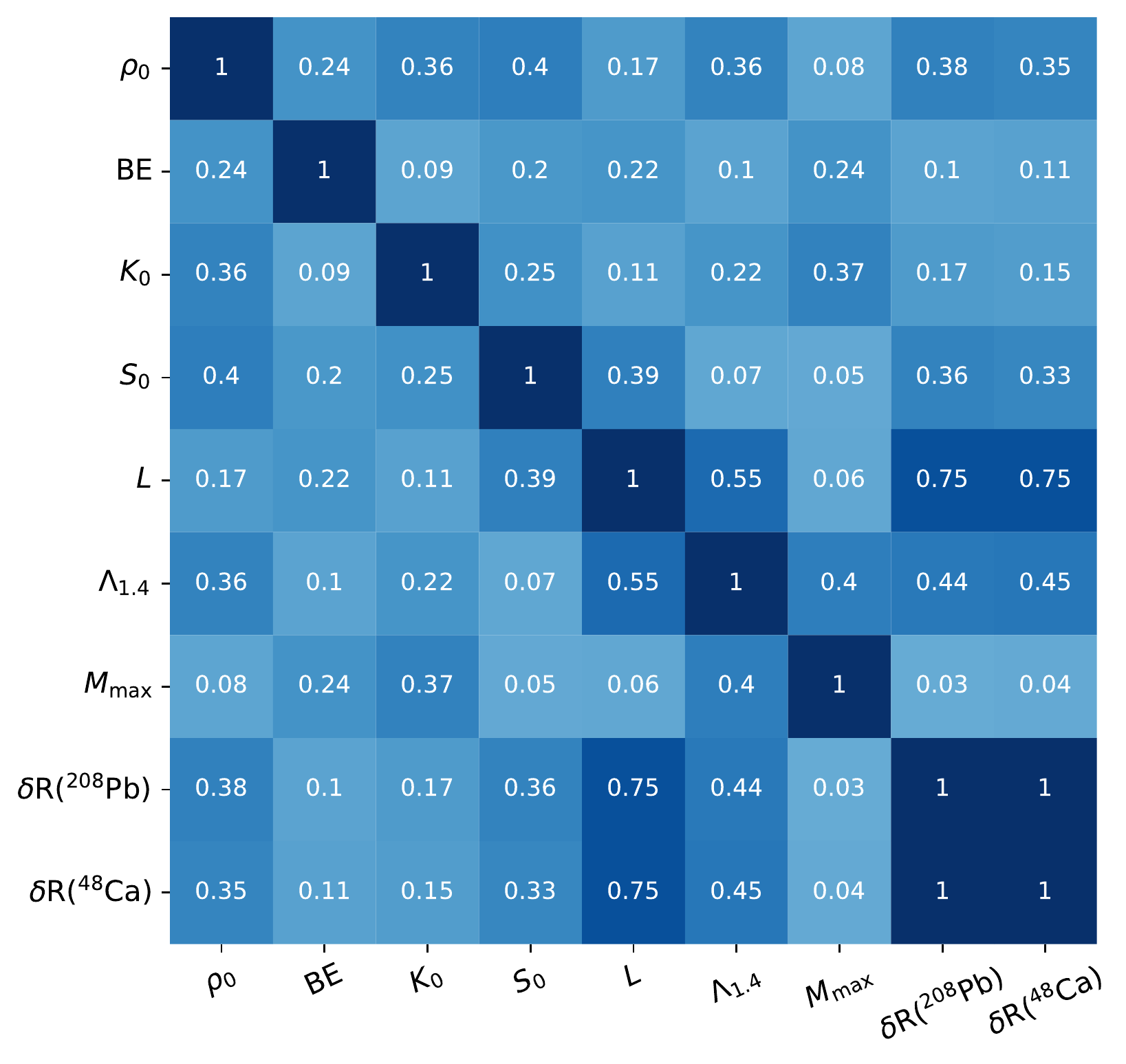}}
\caption{
The correlation matrix between nuclear saturation properties and NS properties for the EoS ensemble shown in Table~\ref{t:sat}. }
\label{f:fig4}
\end{figure}

Finally, in Fig.~\ref{f:fig4}, we display a correlation matrix among the saturation properties shown in Table~\ref{t:sat}, with $\Lambda_{1.4}$, $M_{\rm max}$ and the neutron skin thickness for $^{48}$Ca and $^{208}$Pb. The matrix confirms the weak correlation of $\delta R$ with the $L$ parameter already shown in Fig.~\ref{f:fig3}. A weaker degree of correlation is found between $\Lambda_{1.4}$ and $L$ (r=0.55), whereas no evident correlation between $\Lambda_{1.4}$ and $S_0$ and $K_0$ is found. 

\section{Conclusions}
\label{sec:5}

In this work, we have analyzed the predictions of microscopic and phenomenological EoS for the neutron skin thickness $\delta R$ of $^{48}$Ca and $^{208}$Pb, and compared with the recent experimental data, CREX and PREX I+II. We have used an ensemble of different EoS models, that includes microscopic calculations based on the (Dirac)Brueckner--Hartree--Fock  theory, the variational method, and Quantum Monte Carlo techniques, as well as several phenomenological Skyrme, and RMF models. The chosen EoSs are compatible with the constraints imposed by laboratory data on saturation properties of finite nuclei, and observational data by NICER and GW170817 regarding the NS mass and $\Lambda_{1.4}$.

We have found a linear correlation between the neutron skin thickness $\delta R$ of $^{48}$Ca and $^{208}$Pb and the $L$ parameter, as already pointed out by several authors using nonrelativistic and relativistic phenomenological models. A weaker linear correlation with the tidal deformability $\Lambda_{1.4}$ is evident.

The most important result of our analysis is that the same EoS cannot reproduce at the same time the CREX and PREX I+II experimental data. Therefore, those measurements do not allow us to select the most compatible EoS among the ones  considered in this work. Future NS observations, along with planned experiments in existing facilities or in next-generation radioactive ion beam laboratories, are fundamental to provide more stringent constraints on the nuclear EoS, thus finally improving our knowledge of the extreme density matter conditions.

\section*{Author Contributions} 
{Conceptualization, G.F.B., I.V.; methodology, G.F.B., I.V.;
software, H.C.D., I.V.; validation, G.F.B, H.C.D and I.V.;
writing—original draft preparation, G.F.B.; writing—review and editing, I.V., H.C.D.; visualization, H.C.D..}

\vspace{6pt} 


\newcommand{\apjl}{Astrophys. J. Lett.\ }
\newcommand{\apj}{Astrophys. J.\ }
\newcommand{\physrep}{Phys. Rep.\ }
\newcommand{\mnras}{Mon. Not. R. Astron. Soc.\ }
\newcommand{\aap}{Astron. Astrophys.\ }
\newcommand{\prc}{Phys. Rev. C\ }
\newcommand{\prd}{Phys. Rev. D\ }
\newcommand{\prl}{Phys. Rev. Lett.\ }
\newcommand{\nphysa}{Nucl. Phys. A\ }
\newcommand{\plb}{Phys. Lett. B\ }
\newcommand{\epja}{EPJA\ }


\bibliographystyle{Definitions/Frontiers-Harvard} 
\bibliography{main}

\begin{thebibliography}{106}
\providecommand{\natexlab}[1]{#1}
\expandafter\ifx\csname urlstyle\endcsname\relax
  \providecommand{\doi}[1]{doi:\discretionary{}{}{}#1}\else
  \providecommand{\doi}{doi:\discretionary{}{}{}\begingroup
  \urlstyle{rm}\Url}\fi
\providecommand{\selectlanguage}[1]{\relax}
\providecommand{\bibAnnoteFile}[1]{%
  \IfFileExists{#1}{\begin{quotation}\noindent\textsc{Key:} #1\\
  \textsc{Annotation:}\ \input{#1}\end{quotation}}{}}
\providecommand{\bibAnnote}[2]{%
  \begin{quotation}\noindent\textsc{Key:} #1\\
  \textsc{Annotation:}\ #2\end{quotation}}

\bibitem[{Abbott et~al.(2017)}]{merger}
Abbott, B. et~al. (2017).
\newblock {GW170817: Observation of Gravitational Waves from a Binary Neutron
  Star Inspiral}.
\newblock \emph{Phys. Rev. Lett.} 119, 161101.
\newblock \doi{10.1103/PhysRevLett.119.161101}
\bibAnnoteFile{merger}

\bibitem[{Abbott et~al.(2018)}]{mergerl}
Abbott, B.~P. et~al. (2018).
\newblock {GW170817: Measurements of neutron star radii and equation of state}.
\newblock \emph{Phys. Rev. Lett.} 121, 161101.
\newblock \doi{10.1103/PhysRevLett.121.161101}
\bibAnnoteFile{mergerl}

\bibitem[{Abbott et~al.(2019)}]{mergerx}
Abbott, B.~P. et~al. (2019).
\newblock Properties of the binary neutron star merger {GW}170817.
\newblock \emph{Phys. Rev. X} 9, 011001.
\newblock \doi{10.1103/PhysRevX.9.011001}
\bibAnnoteFile{mergerx}

\bibitem[{Adhikari et~al.(2021)}]{Adhikari_2021}
Adhikari, D. et~al. (2021).
\newblock Accurate determination of the neutron skin thickness of
  $^{208}\mathrm{Pb}$ through parity-violation in electron scattering.
\newblock \emph{Phys. Rev. Lett.} 126, 172502.
\newblock \doi{10.1103/PhysRevLett.126.172502}
\bibAnnoteFile{Adhikari_2021}

\bibitem[{Adhikari et~al.(2022)}]{Adhikari_2022}
Adhikari, D. et~al. (2022).
\newblock Precision determination of the neutral weak form factor of
  $^{48}\mathrm{Ca}$.
\newblock \emph{Phys. Rev. Lett.} 129, 042501.
\newblock \doi{10.1103/PhysRevLett.129.042501}
\bibAnnoteFile{Adhikari_2022}

\bibitem[{{Akmal} et~al.(1998){Akmal}, {Pandharipande}, and
  {Ravenhall}}]{apr1998}
{Akmal}, A., {Pandharipande}, V.~R., and {Ravenhall}, D.~G. (1998).
\newblock {Equation of state of nucleon matter and neutron star structure}.
\newblock \emph{\prc} 58, 1804--1828.
\newblock \doi{10.1103/PhysRevC.58.1804}
\bibAnnoteFile{apr1998}

\bibitem[{Antoniadis et~al.(2013)}]{heavy2}
Antoniadis, J. et~al. (2013).
\newblock {A Massive Pulsar in a Compact Relativistic Binary}.
\newblock \emph{Science} 340, 6131.
\newblock \doi{10.1126/science.1233232}
\bibAnnoteFile{heavy2}

\bibitem[{{Audi} et~al.(2003){Audi}, {Wapstra}, and {Thibault}}]{audi03}
{Audi}, G., {Wapstra}, A.~H., and {Thibault}, C. (2003).
\newblock {The AME2003 atomic mass evaluation . (II). Tables, graphs and
  references}.
\newblock \emph{\nphysa} 729, 337--676.
\newblock \doi{10.1016/j.nuclphysa.2003.11.003}
\bibAnnoteFile{audi03}

\bibitem[{{Baldo}(1999)}]{baldo1999}
{Baldo}, M. (1999).
\newblock {Nuclear Methods And The Nuclear Equation Of State}.
\newblock \emph{International Review of Nuclear Physics (World Scientific,
  Singapore)} 8.
\newblock \doi{10.1142/2657}
\bibAnnoteFile{baldo1999}

\bibitem[{{Baldo} et~al.(1997){Baldo}, {Bombaci}, and {Burgio}}]{uix3}
{Baldo}, M., {Bombaci}, I., and {Burgio}, G.~F. (1997).
\newblock {Microscopic nuclear equation of state with three-body forces and
  neutron star structure}.
\newblock \emph{\aap} 328, 274--282.
\newblock \doi{1997A&A...328..274B}
\bibAnnoteFile{uix3}

\bibitem[{{Baldo} and {Burgio}(2016)}]{2016PrPNP}
{Baldo}, M. and {Burgio}, G.~F. (2016).
\newblock {The nuclear symmetry energy}.
\newblock \emph{Progress in Particle and Nuclear Physics} 91, 203--258.
\newblock \doi{10.1016/j.ppnp.2016.06.006}
\bibAnnoteFile{2016PrPNP}

\bibitem[{{Baldo} and {Fukukawa}(2014)}]{2014PhRvL.113x2501B}
{Baldo}, M. and {Fukukawa}, K. (2014).
\newblock {Nuclear Matter from Effective Quark-Quark Interaction}.
\newblock \emph{\prl} 113, 242501.
\newblock \doi{10.1103/PhysRevLett.113.242501}
\bibAnnoteFile{2014PhRvL.113x2501B}

\bibitem[{Baym et~al.(1971)Baym, Pethick, and Sutherland}]{bps}
Baym, G., Pethick, C., and Sutherland, P. (1971).
\newblock {The Ground state of matter at high densities: Equation of state and
  stellar models}.
\newblock \emph{Astrophys. J.} 170, 299--317.
\newblock \doi{10.1086/151216}
\bibAnnoteFile{bps}

\bibitem[{Benhar and Lovato(2017)}]{Benhar_2017}
Benhar, O. and Lovato, A. (2017).
\newblock Perturbation theory of nuclear matter with a microscopic effective
  interaction.
\newblock \emph{Phys. Rev. C} 96, 054301.
\newblock \doi{10.1103/PhysRevC.96.054301}
\bibAnnoteFile{Benhar_2017}

\bibitem[{{Boguta} and {Bodmer}(1977)}]{rmfbb}
{Boguta}, J. and {Bodmer}, A.~R. (1977).
\newblock {Relativistic calculation of nuclear matter and the nuclear surface}.
\newblock \emph{\nphysa} 292, 413--428.
\newblock \doi{10.1016/0375-9474(77)90626-1}
\bibAnnoteFile{rmfbb}

\bibitem[{{Brown}(2000)}]{skin1}
{Brown}, B.~A. (2000).
\newblock {Neutron Radii in Nuclei and the Neutron Equation of State}.
\newblock \emph{\prl} 85, 5296--5299.
\newblock \doi{10.1103/PhysRevLett.85.5296}
\bibAnnoteFile{skin1}

\bibitem[{Burgio et~al.(2021)Burgio, Schulze, Vidaña, and Wei}]{Burgio_2021}
Burgio, G., Schulze, H.-J., Vidaña, I., and Wei, J.-B. (2021).
\newblock Neutron stars and the nuclear equation of state.
\newblock \emph{Progress in Particle and Nuclear Physics} 120, 103879
\bibAnnoteFile{Burgio_2021}

\bibitem[{{Burgio} and {Fantina}(2018)}]{2018Chap6}
{Burgio}, G.~F. and {Fantina}, A.~F. (2018).
\newblock {Nuclear Equation of state for Compact Stars and Supernovae}.
\newblock \emph{Astrophys. Space Sci.Libr.} 457, 255
\bibAnnoteFile{2018Chap6}

\bibitem[{Burgio and Vidaña(2020)}]{universe6080119}
Burgio, G.~F. and Vidaña, I. (2020).
\newblock The equation of state of nuclear matter: From finite nuclei to
  neutron stars.
\newblock \emph{Universe} 6, 119.
\newblock \doi{10.3390/universe6080119}
\bibAnnoteFile{universe6080119}

\bibitem[{{Carbone} et~al.(2010){Carbone}, {Col{\`o}}, {Bracco}, {Cao},
  {Bortignon}, {Camera} et~al.}]{pygmy2}
{Carbone}, A., {Col{\`o}}, G., {Bracco}, A., {Cao}, L.-G., {Bortignon}, P.~F.,
  {Camera}, F., et~al. (2010).
\newblock {Constraints on the symmetry energy and neutron skins from pygmy
  resonances in Ni68 and Sn132}.
\newblock \emph{\prc} 81, 041301.
\newblock \doi{10.1103/PhysRevC.81.041301}
\bibAnnoteFile{pygmy2}

\bibitem[{{Centelles} et~al.(2009){Centelles}, {Roca-Maza}, {Vi{\~n}as}, and
  {Warda}}]{cen09}
{Centelles}, M., {Roca-Maza}, X., {Vi{\~n}as}, X., and {Warda}, M. (2009).
\newblock {Nuclear Symmetry Energy Probed by Neutron Skin Thickness of Nuclei}.
\newblock \emph{\prl} 102, 122502.
\newblock \doi{10.1103/PhysRevLett.102.122502}
\bibAnnoteFile{cen09}

\bibitem[{{Chabanat}(1995)}]{sly}
{Chabanat}, E. (1995).
\newblock {Ph.D. Thesis, Universit\`e Claude Bernard Lyon-1}.
\newblock \emph{Report No. LYCENT 9501 (unpublished)}
\bibAnnoteFile{sly}

\bibitem[{{Chabanat} et~al.(1997){Chabanat}, {Bonche}, {Haensel}, {Meyer}, and
  {Schaeffer}}]{sly230a1}
{Chabanat}, E., {Bonche}, P., {Haensel}, P., {Meyer}, J., and {Schaeffer}, R.
  (1997).
\newblock {A Skyrme parametrization from subnuclear to neutron star densities}.
\newblock \emph{\nphysa} 627, 710--746.
\newblock \doi{10.1016/S0375-9474(97)00596-4}
\bibAnnoteFile{sly230a1}

\bibitem[{{Chabanat} et~al.(1998){Chabanat}, {Bonche}, {Haensel}, {Meyer}, and
  {Schaeffer}}]{sly230a2}
{Chabanat}, E., {Bonche}, P., {Haensel}, P., {Meyer}, J., and {Schaeffer}, R.
  (1998).
\newblock {A Skyrme parametrization from subnuclear to neutron star
  densitiesPart II. Nuclei far from stabilities}.
\newblock \emph{\nphysa} 635, 231--256.
\newblock \doi{10.1016/S0375-9474(98)00180-8}
\bibAnnoteFile{sly230a2}

\bibitem[{{Chen} et~al.(2010){Chen}, {Ko}, {Li}, and {Xu}}]{chen2010}
{Chen}, L.-W., {Ko}, C.~M., {Li}, B.-A., and {Xu}, J. (2010).
\newblock {Density slope of the nuclear symmetry energy from the neutron skin
  thickness of heavy nuclei}.
\newblock \emph{\prc} 82, 024321.
\newblock \doi{10.1103/PhysRevC.82.024321}
\bibAnnoteFile{chen2010}

\bibitem[{Chen and Piekarewicz(2015)}]{FSUGarnet}
Chen, W.-C. and Piekarewicz, J. (2015).
\newblock {Searching for isovector signatures in the neutron-rich oxygen and
  calcium isotopes}.
\newblock \emph{Phys. Lett. B} 748, 284--288.
\newblock \doi{10.1016/j.physletb.2015.07.020}
\bibAnnoteFile{FSUGarnet}

\bibitem[{{Col{\`o}} et~al.(2004){Col{\`o}}, {van Giai}, {Meyer}, {Bennaceur},
  and {Bonche}}]{colo04}
{Col{\`o}}, G., {van Giai}, N., {Meyer}, J., {Bennaceur}, K., and {Bonche}, P.
  (2004).
\newblock {Microscopic determination of the nuclear incompressibility within
  the nonrelativistic framework}.
\newblock \emph{\prc} 70, 024307.
\newblock \doi{10.1103/PhysRevC.70.024307}
\bibAnnoteFile{colo04}

\bibitem[{{Cromartie} et~al.(2019)}]{cromartie}
{Cromartie}, H.~T. et~al. (2019).
\newblock Relativistic shapiro delay measurements of an extremely massive
  millisecond pulsar.
\newblock \emph{Nature Astronomy} 4, 72–76.
\newblock \doi{10.1038/s41550-019-0880-2}
\bibAnnoteFile{cromartie}

\bibitem[{{Danielewicz} and {Lee}(2014)}]{danlee2014}
{Danielewicz}, P. and {Lee}, J. (2014).
\newblock {Symmetry energy II: Isobaric analog states}.
\newblock \emph{\nphysa} 922, 1--70.
\newblock \doi{10.1016/j.nuclphysa.2013.11.005}
\bibAnnoteFile{danlee2014}

\bibitem[{Das et~al.(2021)Das, Kumar, Kumar, Biswal, and Patra}]{BigApple_B}
Das, H.~C., Kumar, A., Kumar, B., Biswal, S.~K., and Patra, S.~K. (2021).
\newblock Big{A}pple force and its implications to finite nuclei and
  astrophysical objects.
\newblock \emph{International Journal of Modern Physics E} 30, 2150088.
\newblock \doi{10.1142/S0218301321500889}
\bibAnnoteFile{BigApple_B}

\bibitem[{{de Vries} et~al.(1987){de Vries}, {de Jager}, and {de
  Vries}}]{vries87}
{de Vries}, H., {de Jager}, C.~W., and {de Vries}, C. (1987).
\newblock {Nuclear Charge-Density-Distribution Parameters from Electron
  Scattering}.
\newblock \emph{Atomic Data and Nuclear Data Tables} 36, 495.
\newblock \doi{10.1016/0092-640X(87)90013-1}
\bibAnnoteFile{vries87}

\bibitem[{Demorest et~al.(2010)Demorest, Pennucci, Ransom, Roberts, and
  Hessels}]{demorest2010}
Demorest, P.~B., Pennucci, T., Ransom, S.~M., Roberts, M.~S., and Hessels,
  J.~W. (2010).
\newblock A two-solar-mass neutron star measured using shapiro delay.
\newblock \emph{Nature} 467, 1081--3.
\newblock \doi{10.1038/nature09466}
\bibAnnoteFile{demorest2010}

\bibitem[{Dhiman et~al.(2007)Dhiman, Kumar, and Agrawal}]{BSR}
Dhiman, S.~K., Kumar, R., and Agrawal, B.~K. (2007).
\newblock Nonrotating and rotating neutron stars in the extended field
  theoretical model.
\newblock \emph{Phys. Rev. C} 76, 045801.
\newblock \doi{10.1103/PhysRevC.76.045801}
\bibAnnoteFile{BSR}

\bibitem[{{Essick} et~al.(2021){Essick}, {Tews}, {Landry}, and
  {Schwenk}}]{Essick21}
{Essick}, R., {Tews}, I., {Landry}, P., and {Schwenk}, A. (2021).
\newblock {Astrophysical Constraints on the Symmetry Energy and the Neutron
  Skin of $^{208}$Pb with Minimal Modeling Assumptions}.
\newblock \emph{\prl} 127, 192701.
\newblock \doi{10.1103/PhysRevLett.127.192701}
\bibAnnoteFile{Essick21}

\bibitem[{Fattoyev et~al.(2020)Fattoyev, Horowitz, Piekarewicz, and
  Reed}]{BigApple_A}
Fattoyev, F.~J., Horowitz, C.~J., Piekarewicz, J., and Reed, B. (2020).
\newblock G{W}190814: Impact of a 2.6 solar mass neutron star on the nucleonic
  equations of state.
\newblock \emph{Phys. Rev. C} 102, 065805.
\newblock \doi{10.1103/PhysRevC.102.065805}
\bibAnnoteFile{BigApple_A}

\bibitem[{Fattoyev and Piekarewicz(2010)}]{IUFSU}
Fattoyev, F.~J. and Piekarewicz, J. (2010).
\newblock Relativistic models of the neutron-star matter equation of state.
\newblock \emph{Phys. Rev. C} 82, 025805.
\newblock \doi{10.1103/PhysRevC.82.025805}
\bibAnnoteFile{IUFSU}

\bibitem[{Feynman et~al.(1949)Feynman, Metropolis, and Teller}]{fey}
Feynman, R.~P., Metropolis, N., and Teller, E. (1949).
\newblock {Equations of State of Elements Based on the Generalized Fermi-Thomas
  Theory}.
\newblock \emph{Phys. Rev.} 75, 1561--1573.
\newblock \doi{10.1103/PhysRev.75.1561}
\bibAnnoteFile{fey}

\bibitem[{Flanagan and Hinderer(2008)}]{flan}
Flanagan, E.~E. and Hinderer, T. (2008).
\newblock {Constraining neutron star tidal Love numbers with gravitational wave
  detectors}.
\newblock \emph{Phys. Rev. D} 77, 021502.
\newblock \doi{10.1103/PhysRevD.77.021502}
\bibAnnoteFile{flan}

\bibitem[{Fonseca et~al.(2016)}]{fonseca2016}
Fonseca, E. et~al. (2016).
\newblock {The NANOGrav Nine-year Data Set: Mass and Geometric Measurements of
  Binary Millisecond Pulsars}.
\newblock \emph{Astrophys. J.} 832, 167.
\newblock \doi{10.3847/0004-637X/832/2/167}
\bibAnnoteFile{fonseca2016}

\bibitem[{{Fuchs} et~al.(2001){Fuchs}, {Faessler}, {Zabrodin}, and
  {Zheng}}]{fuchs01}
{Fuchs}, C., {Faessler}, A., {Zabrodin}, E., and {Zheng}, Y.-M. (2001).
\newblock {Probing the Nuclear Equation of State by K$^{+}$ Production in
  Heavy-Ion Collisions}.
\newblock \emph{\prl} 86, 1974--1977.
\newblock \doi{10.1103/PhysRevLett.86.1974}
\bibAnnoteFile{fuchs01}

\bibitem[{{Fukukawa} et~al.(2015){Fukukawa}, {Baldo}, {Burgio}, {Lo Monaco},
  and {Schulze}}]{2015PhRvC..92f5802F}
{Fukukawa}, K., {Baldo}, M., {Burgio}, G.~F., {Lo Monaco}, L., and {Schulze},
  H.-J. (2015).
\newblock {Nuclear matter equation of state from a quark-model nucleon-nucleon
  interaction}.
\newblock \emph{\prc} 92, 065802.
\newblock \doi{10.1103/PhysRevC.92.065802}
\bibAnnoteFile{2015PhRvC..92f5802F}

\bibitem[{{Furnstahl}(2002)}]{fur02}
{Furnstahl}, R.~J. (2002).
\newblock {Neutron radii in mean-field models}.
\newblock \emph{\nphysa} 706, 85--110.
\newblock \doi{10.1016/S0375-9474(02)00867-9}
\bibAnnoteFile{fur02}

\bibitem[{{Gandolfi} et~al.(2010){Gandolfi}, {Illarionov}, {Fantoni}, {Miller},
  {Pederiva}, and {Schmidt}}]{afdmc}
{Gandolfi}, S., {Illarionov}, A.~Y., {Fantoni}, S., {Miller}, J.~C.,
  {Pederiva}, F., and {Schmidt}, K.~E. (2010).
\newblock {Microscopic calculation of the equation of state of nuclear matter
  and neutron star structure}.
\newblock \emph{Mon. Not. Roy. Astron. Soc.} 404, L35--L39.
\newblock \doi{10.1111/j.1745-3933.2010.00829.x}
\bibAnnoteFile{afdmc}

\bibitem[{Goriely et~al.(2013)Goriely, Chamel, and Pearson}]{Goriely2013}
Goriely, S., Chamel, N., and Pearson, J.~M. (2013).
\newblock Further explorations of skyrme-hartree-fock-bogoliubov mass formulas.
  xiii. the 2012 atomic mass evaluation and the symmetry coefficient.
\newblock \emph{Phys. Rev. C} 88, 024308.
\newblock \doi{10.1103/PhysRevC.88.024308}
\bibAnnoteFile{Goriely2013}

\bibitem[{{Grang{\'e}} et~al.(1989){Grang{\'e}}, {Lejeune}, {Martzolff}, and
  {Mathiot}}]{glmm}
{Grang{\'e}}, P., {Lejeune}, A., {Martzolff}, M., and {Mathiot}, J.-F. (1989).
\newblock {Consistent three-nucleon forces in the nuclear many-body problem}.
\newblock \emph{\prc} 40, 1040--1060.
\newblock \doi{10.1103/PhysRevC.40.1040}
\bibAnnoteFile{glmm}

\bibitem[{{Gross-Boelting} et~al.(1999){Gross-Boelting}, {Fuchs}, and
  {Faessler}}]{dbhf3}
{Gross-Boelting}, T., {Fuchs}, C., and {Faessler}, A. (1999).
\newblock {Covariant representations of the relativistic Brueckner T-matrix and
  the nuclear matter problem}.
\newblock \emph{Nucl. Phys. A} 648, 105--137.
\newblock \doi{10.1016/S0375-9474(99)00022-6}
\bibAnnoteFile{dbhf3}

\bibitem[{Guillot et~al.(2013)Guillot, Servillat, Webb, and Rutledge}]{gui2013}
Guillot, S., Servillat, M., Webb, N.~A., and Rutledge, R.~E. (2013).
\newblock Measurement of the radius of neutron stars with high signal-to-noise
  quiescent low-mass x-ray binaries in globular clusters.
\newblock \emph{Astrophys. J.} 772, 7.
\newblock \doi{10.1088/0004-637X/772/1/7}
\bibAnnoteFile{gui2013}

\bibitem[{Hartle(1967)}]{hartle}
Hartle, J.~B. (1967).
\newblock {Slowly rotating relativistic stars. 1. Equations of structure}.
\newblock \emph{Astrophys. J.} 150, 1005--1029.
\newblock \doi{10.1086/149400}
\bibAnnoteFile{hartle}

\bibitem[{Hinderer(2008)}]{hinder2008}
Hinderer, T. (2008).
\newblock {Tidal Love numbers of neutron stars}.
\newblock \emph{Astrophys. J.} 677, 1216--1220.
\newblock \doi{10.1086/533487}
\bibAnnoteFile{hinder2008}

\bibitem[{Hinderer(2009)}]{hinder2009}
Hinderer, T. (2009).
\newblock Erratum: "{T}idal {L}ove numbers of neutron stars" (2008, apj, 677,
  1216).
\newblock \emph{Astrophys. J.} 697, 964.
\newblock \doi{10.1088/0004-637X/697/1/964}
\bibAnnoteFile{hinder2009}

\bibitem[{Hinderer et~al.(2010)Hinderer, Lackey, Lang, and Read}]{hinder2010}
Hinderer, T., Lackey, B.~D., Lang, R.~N., and Read, J.~S. (2010).
\newblock {Tidal deformability of neutron stars with realistic equations of
  state and their gravitational wave signatures in binary inspiral}.
\newblock \emph{Phys. Rev. D} 81, 123016.
\newblock \doi{10.1103/PhysRevD.81.123016}
\bibAnnoteFile{hinder2010}

\bibitem[{{Horowitz} and {Piekarewicz}(2001)}]{horo01}
{Horowitz}, C.~J. and {Piekarewicz}, J. (2001).
\newblock {Neutron Star Structure and the Neutron Radius of $^{208}$Pb}.
\newblock \emph{\prl} 86, 5647--5650.
\newblock \doi{10.1103/PhysRevLett.86.5647}
\bibAnnoteFile{horo01}

\bibitem[{{Horowitz} et~al.(2001){Horowitz}, {Pollock}, {Souder}, and
  {Michaels}}]{skin3}
{Horowitz}, C.~J., {Pollock}, S.~J., {Souder}, P.~A., and {Michaels}, R.
  (2001).
\newblock {Parity violating measurements of neutron densities}.
\newblock \emph{\prc} 63, 025501.
\newblock \doi{10.1103/PhysRevC.63.025501}
\bibAnnoteFile{skin3}

\bibitem[{{Klimkiewicz} et~al.(2007){Klimkiewicz}, {Paar}, {Adrich}, {Fallot},
  {Boretzky}, {Aumann} et~al.}]{pygmy1}
{Klimkiewicz}, A., {Paar}, N., {Adrich}, P., {Fallot}, M., {Boretzky}, K.,
  {Aumann}, T., et~al. (2007).
\newblock {Nuclear symmetry energy and neutron skins derived from pygmy dipole
  resonances}.
\newblock \emph{\prc} 76, 051603.
\newblock \doi{10.1103/PhysRevC.76.051603}
\bibAnnoteFile{pygmy1}

\bibitem[{Kumar et~al.(2018)Kumar, Patra, and Agrawal}]{IOPB}
Kumar, B., Patra, S.~K., and Agrawal, B.~K. (2018).
\newblock New relativistic effective interaction for finite nuclei, infinite
  nuclear matter, and neutron stars.
\newblock \emph{Phys. Rev. C} 97, 045806.
\newblock \doi{10.1103/PhysRevC.97.045806}
\bibAnnoteFile{IOPB}

\bibitem[{Kumar et~al.(2017)Kumar, Singh, Agrawal, and Patra}]{G3}
Kumar, B., Singh, S.~K., Agrawal, B.~K., and Patra, S.~K. (2017).
\newblock {New parameterization of the effective field theory motivated
  relativistic mean field model}.
\newblock \emph{Nucl. Phys. A} 966, 197--207.
\newblock \doi{10.1016/j.nuclphysa.2017.07.001}
\bibAnnoteFile{G3}

\bibitem[{Kumar et~al.(2006)Kumar, Agrawal, and Dhiman}]{FSUGZ03}
Kumar, R., Agrawal, B.~K., and Dhiman, S.~K. (2006).
\newblock Effects of \ensuremath{\omega} meson self-coupling on the properties
  of finite nuclei and neutron stars.
\newblock \emph{Phys. Rev. C} 74, 034323.
\newblock \doi{10.1103/PhysRevC.74.034323}
\bibAnnoteFile{FSUGZ03}

\bibitem[{Lattimer(2012)}]{mass}
Lattimer, J.~M. (2012).
\newblock The nuclear equation of state and neutron star masses.
\newblock \emph{Ann. Rev. Nucl. Sci.} 62, 485--515.
\newblock \doi{10.1146/annurev-nucl-102711-095018}
\bibAnnoteFile{mass}

\bibitem[{Lattimer(2023)}]{Lattimer_2023}
Lattimer, J.~M. (2023).
\newblock Constraints on nuclear symmetry energy parameters.
\newblock \emph{Particles} 6, 30–56.
\newblock \doi{10.3390/particles6010003}
\bibAnnoteFile{Lattimer_2023}

\bibitem[{{Li} et~al.(2008){Li}, {Lombardo}, {Schulze}, and {Zuo}}]{tbfnij}
{Li}, Z.~H., {Lombardo}, U., {Schulze}, H.-J., and {Zuo}, W. (2008).
\newblock {Consistent nucleon-nucleon potentials and three-body forces}.
\newblock \emph{\prc} 77, 034316.
\newblock \doi{10.1103/PhysRevC.77.034316}
\bibAnnoteFile{tbfnij}

\bibitem[{Li and Schulze(2008)}]{li08}
Li, Z.~H. and Schulze, H.~J. (2008).
\newblock {Neutron star structure with modern nucleonic three-body forces}.
\newblock \emph{Phys. Rev. C} 78, 028801.
\newblock \doi{10.1103/PhysRevC.78.028801}
\bibAnnoteFile{li08}

\bibitem[{{Li, Bao-An} et~al.(2014){Li, Bao-An}, {Ramos, À.}, {Verde, G.}, and
  {Vidaña, I.}}]{EPJA_2014}
{Li, Bao-An}, {Ramos, À.}, {Verde, G.}, and {Vidaña, I.} (2014).
\newblock Topical issue on nuclear symmetry energy.
\newblock \emph{Eur. Phys. J. A} 50, 9.
\newblock \doi{10.1140/epja/i2014-14009-x}
\bibAnnoteFile{EPJA_2014}

\bibitem[{Machleidt(1989)}]{bonn2}
Machleidt, R. (1989).
\newblock {The Meson theory of nuclear forces and nuclear structure}.
\newblock \emph{Adv. Nucl. Phys.} 19, 189--376.
\newblock \doi{10.1007/978-1-4613-9907-0}
\bibAnnoteFile{bonn2}

\bibitem[{Machleidt et~al.(1987)Machleidt, Holinde, and Elster}]{bonn1}
Machleidt, R., Holinde, K., and Elster, C. (1987).
\newblock {The Bonn Meson Exchange Model for the Nucleon Nucleon Interaction}.
\newblock \emph{\physrep} 149, 1--89.
\newblock \doi{10.1016/S0370-1573(87)80002-9}
\bibAnnoteFile{bonn1}

\bibitem[{{Margueron} et~al.(2018){Margueron}, {Hoffmann Casali}, and
  {Gulminelli}}]{margue2018a}
{Margueron}, J., {Hoffmann Casali}, R., and {Gulminelli}, F. (2018).
\newblock {Equation of state for dense nucleonic matter from metamodeling. I.
  Foundational aspects}.
\newblock \emph{\prc} 97, 025805.
\newblock \doi{10.1103/PhysRevC.97.025805}
\bibAnnoteFile{margue2018a}

\bibitem[{{Miller} et~al.(2019)}]{millernicer}
{Miller}, M.~C. et~al. (2019).
\newblock {PSR J0030+0451 Mass and Radius from NICER Data and Implications for
  the Properties of Neutron Star Matter}.
\newblock \emph{Astrophys. J. Lett.} 887, L24.
\newblock \doi{10.3847/2041-8213/ab50c5}
\bibAnnoteFile{millernicer}

\bibitem[{{M{\"o}ller} et~al.(2012){M{\"o}ller}, {Myers}, {Sagawa}, and
  {Yoshida}}]{moller2012}
{M{\"o}ller}, P., {Myers}, W.~D., {Sagawa}, H., and {Yoshida}, S. (2012).
\newblock {New Finite-Range Droplet Mass Model and Equation-of-State
  Parameters}.
\newblock \emph{\prl} 108, 052501.
\newblock \doi{10.1103/PhysRevLett.108.052501}
\bibAnnoteFile{moller2012}

\bibitem[{Mondal et~al.(2016)Mondal, Agrawal, De, and Samaddar}]{SINPA_N_B}
Mondal, C., Agrawal, B.~K., De, J.~N., and Samaddar, S.~K. (2016).
\newblock Sensitivity of elements of the symmetry energy of nuclear matter to
  the properties of neutron-rich systems.
\newblock \emph{Phys. Rev. C} 93, 044328.
\newblock \doi{10.1103/PhysRevC.93.044328}
\bibAnnoteFile{SINPA_N_B}

\bibitem[{Nagels et~al.(1978)Nagels, Rijken, and de~Swart}]{nij1}
Nagels, M.~M., Rijken, T.~A., and de~Swart, J.~J. (1978).
\newblock {A Low-Energy Nucleon-Nucleon Potential from Regge Pole Theory}.
\newblock \emph{Phys. Rev. D} 17, 768.
\newblock \doi{10.1103/PhysRevD.17.768}
\bibAnnoteFile{nij1}

\bibitem[{Negele and Vautherin(1973)}]{NV}
Negele, J.~W. and Vautherin, D. (1973).
\newblock {Neutron star matter at subnuclear densities}.
\newblock \emph{Nucl. Phys. A} 207, 298--320.
\newblock \doi{10.1016/0375-9474(73)90349-7}
\bibAnnoteFile{NV}

\bibitem[{{Nik{\v{s}}i{\'c}} et~al.(2002){Nik{\v{s}}i{\'c}}, {Vretenar},
  {Finelli}, and {Ring}}]{ddme}
{Nik{\v{s}}i{\'c}}, T., {Vretenar}, D., {Finelli}, P., and {Ring}, P. (2002).
\newblock {Relativistic Hartree-Bogoliubov model with density-dependent
  meson-nucleon couplings}.
\newblock \emph{\prc} 66, 024306.
\newblock \doi{10.1103/PhysRevC.66.024306}
\bibAnnoteFile{ddme}

\bibitem[{{Oertel} et~al.(2017){Oertel}, {Hempel}, {Kl{\"a}hn}, and
  {Typel}}]{oertel}
{Oertel}, M., {Hempel}, M., {Kl{\"a}hn}, T., and {Typel}, S. (2017).
\newblock {Equations of state for supernovae and compact stars}.
\newblock \emph{Reviews of Modern Physics} 89, 015007.
\newblock \doi{10.1103/RevModPhys.89.015007}
\bibAnnoteFile{oertel}

\bibitem[{{\"O}zel and Freire(2016)}]{ozel16}
{\"O}zel, F. and Freire, P. (2016).
\newblock {Masses, Radii, and the Equation of State of Neutron Stars}.
\newblock \emph{Ann. Rev. Astron. Astrophys.} 54, 401--440.
\newblock \doi{10.1146/annurev-astro-081915-023322}
\bibAnnoteFile{ozel16}

\bibitem[{{Piekarewicz}(2004)}]{piekarewicz04}
{Piekarewicz}, J. (2004).
\newblock {Unmasking the nuclear matter equation of state}.
\newblock \emph{\prc} 69, 041301.
\newblock \doi{10.1103/PhysRevC.69.041301}
\bibAnnoteFile{piekarewicz04}

\bibitem[{{Piekarewicz}(2010)}]{pieka}
{Piekarewicz}, J. (2010).
\newblock {Do we understand the incompressibility of neutron-rich matter?}
\newblock \emph{Journal of Physics G Nuclear Physics} 37, 064038.
\newblock \doi{10.1088/0954-3899/37/6/064038}
\bibAnnoteFile{pieka}

\bibitem[{Potekhin et~al.(2013)Potekhin, Fantina, Chamel, Pearson, and
  Goriely}]{Potekhin_2013}
Potekhin, A.~Y., Fantina, A.~F., Chamel, N., Pearson, J.~M., and Goriely, S.
  (2013).
\newblock Analytical representations of unified equations of state for
  neutron-star matter.
\newblock \emph{Astronomy \& Astrophysics} 560.
\newblock \doi{10.1051/0004-6361/201321697}
\bibAnnoteFile{Potekhin_2013}

\bibitem[{Pudliner et~al.(1997)Pudliner, Pandharipande, Carlson, Pieper, and
  Wiringa}]{uix2}
Pudliner, B.~S., Pandharipande, V.~R., Carlson, J., Pieper, S.~C., and Wiringa,
  R.~B. (1997).
\newblock {Quantum Monte Carlo calculations of nuclei with A $\leq$ 7}.
\newblock \emph{Phys. Rev. C} 56, 1720--1750.
\newblock \doi{10.1103/PhysRevC.56.1720}
\bibAnnoteFile{uix2}

\bibitem[{Pudliner et~al.(1995)Pudliner, Pandharipande, Carlson, and
  Wiringa}]{uix1}
Pudliner, B.~S., Pandharipande, V.~R., Carlson, J., and Wiringa, R.~B. (1995).
\newblock {Quantum Monte Carlo calculations of A $\leq$ 6 nuclei}.
\newblock \emph{Phys. Rev. Lett.} 74, 4396--4399.
\newblock \doi{10.1103/PhysRevLett.74.4396}
\bibAnnoteFile{uix1}

\bibitem[{Radice et~al.(2018)Radice, Perego, Zappa, and Bernuzzi}]{radice}
Radice, D., Perego, A., Zappa, F., and Bernuzzi, S. (2018).
\newblock {GW170817: Joint Constraint on the Neutron Star Equation of State
  from Multimessenger Observations}.
\newblock \emph{Astrophys. J.} 852, L29.
\newblock \doi{10.3847/2041-8213/aaa402}
\bibAnnoteFile{radice}

\bibitem[{{Reed, B. T., Fattoyev, F. J., Horowitz, C. J., and Piekarewicz,
  J.}(2021)}]{Reed21}
{Reed, B. T., Fattoyev, F. J., Horowitz, C. J., and Piekarewicz, J.} (2021).
\newblock {Implications of PREX-II on the equation of state of neutron-rich
  matter}.
\newblock \emph{Phys. Rev. Lett.} 126, 172502
\bibAnnoteFile{Reed21}

\bibitem[{{{Riley}, T. E. and others}(2019)}]{nicer3}
{{Riley}, T. E. and others} (2019).
\newblock {A NICER View of PSR J0030+0451: Millisecond Pulsar Parameter
  Estimation}.
\newblock \emph{Astrophys. J. Lett.} 887, L21.
\newblock \doi{10.3847/2041-8213/ab481c}
\bibAnnoteFile{nicer3}

\bibitem[{{Roca-Maza} et~al.(2015){Roca-Maza}, {Vi{\~n}as}, {Centelles},
  {Agrawal}, {Col{\`o}}, {Paar} et~al.}]{rocavin2015}
{Roca-Maza}, X., {Vi{\~n}as}, X., {Centelles}, M., {Agrawal}, B.~K.,
  {Col{\`o}}, G., {Paar}, N., et~al. (2015).
\newblock {Neutron skin thickness from the measured electric dipole
  polarizability in $^{68}$Ni$^{120}$Sn and $^{208}$Pb}.
\newblock \emph{\prc} 92, 064304.
\newblock \doi{10.1103/PhysRevC.92.064304}
\bibAnnoteFile{rocavin2015}

\bibitem[{Romani et~al.(2022)Romani, Kandel, Filippenko, Brink, and
  Zheng}]{Romani_2022}
Romani, R.~W., Kandel, D., Filippenko, A.~V., Brink, T.~G., and Zheng, W.
  (2022).
\newblock {PSR J0952{-}0607: The Fastest and Heaviest Known Galactic Neutron
  Star}.
\newblock \emph{Astrophys. J. Lett.} 934, L17.
\newblock \doi{10.3847/2041-8213/ac8007}
\bibAnnoteFile{Romani_2022}

\bibitem[{Routaray et~al.(2023)Routaray, Mohanty, Das, Ghosh, Kalita, Parmar
  et~al.}]{NITR}
Routaray, P., Mohanty, S.~R., Das, H.~C., Ghosh, S., Kalita, P.~J., Parmar, V.,
  et~al. (2023).
\newblock {Investigating dark matter-admixed neutron stars with NITR equation
  of state in light of PSR J0952-0607}.
\newblock \emph{J. Cosmol. Astropart. Phys.} 2023, 073.
\newblock \doi{10.1088/1475-7516/2023/10/073}
\bibAnnoteFile{NITR}

\bibitem[{Russotto et~al.(2023)Russotto, Cozma, De~Filippo, Le~Fèvre, Leifels,
  and Łukasik}]{Russotto_2023}
Russotto, P., Cozma, M.~D., De~Filippo, E., Le~Fèvre, A., Leifels, Y., and
  Łukasik, J. (2023).
\newblock Studies of the equation-of-state of nuclear matter by heavy-ion
  collisions at intermediate energy in the multi-messenger era: A review
  focused on {GSI} results.
\newblock \emph{La Rivista del Nuovo Cimento} 46, 1–70.
\newblock \doi{10.1007/s40766-023-00039-4}
\bibAnnoteFile{Russotto_2023}

\bibitem[{{Russotto} et~al.(2011)}]{2011PhLB..697..471R}
{Russotto}, P. et~al. (2011).
\newblock {Symmetry energy from elliptic flow in $^{197}$Au + $^{197}$Au}.
\newblock \emph{\plb} 697, 471--476.
\newblock \doi{10.1016/j.physletb.2011.02.033}
\bibAnnoteFile{2011PhLB..697..471R}

\bibitem[{{Russotto} et~al.(2016)}]{2016PhRvC..94c4608R}
{Russotto}, P. et~al. (2016).
\newblock {Results of the ASY-EOS experiment at GSI: The symmetry energy at
  suprasaturation density}.
\newblock \emph{\prc} 94, 034608.
\newblock \doi{10.1103/PhysRevC.94.034608}
\bibAnnoteFile{2016PhRvC..94c4608R}

\bibitem[{Salmi et~al.(2024)}]{Salmi:2024bss}
Salmi, T. et~al. (2024).
\newblock {A NICER View of PSR J1231-1411: A Complex Case}.
\newblock \emph{arXiv:2409.14923}
\bibAnnoteFile{Salmi:2024bss}

\bibitem[{{Shlomo} et~al.(2006){Shlomo}, {Kolomietz}, and {Col{\`o}}}]{shlomo}
{Shlomo}, S., {Kolomietz}, V.~M., and {Col{\`o}}, G. (2006).
\newblock {Deducing the nuclear-matter incompressibility coefficient from data
  on isoscalar compression modes}.
\newblock \emph{\epja} 30, 23--30.
\newblock \doi{10.1140/epja/i2006-10100-3}
\bibAnnoteFile{shlomo}

\bibitem[{{Steiner} et~al.(2005){Steiner}, {Prakash}, {Lattimer}, and
  {Ellis}}]{stei05}
{Steiner}, A.~W., {Prakash}, M., {Lattimer}, J.~M., and {Ellis}, P.~J. (2005).
\newblock {Isospin asymmetry in nuclei and neutron stars}.
\newblock \emph{\physrep} 411, 325--375.
\newblock \doi{10.1016/j.physrep.2005.02.004}
\bibAnnoteFile{stei05}

\bibitem[{Stoks et~al.(1994)Stoks, Klomp, Terheggen, and de~Swart}]{nij2}
Stoks, V. G.~J., Klomp, R. A.~M., Terheggen, C. P.~F., and de~Swart, J.~J.
  (1994).
\newblock {Construction of high quality N N potential models}.
\newblock \emph{Phys. Rev. C} 49, 2950--2962.
\newblock \doi{10.1103/PhysRevC.49.2950}
\bibAnnoteFile{nij2}

\bibitem[{{Stone} and {Reinhard}(2007)}]{stone}
{Stone}, J.~R. and {Reinhard}, P.~G. (2007).
\newblock {The Skyrme interaction in finite nuclei and nuclear matter}.
\newblock \emph{Progress in Particle and Nuclear Physics} 58, 587--657.
\newblock \doi{10.1016/j.ppnp.2006.07.001}
\bibAnnoteFile{stone}

\bibitem[{Sulaksono and Mart(2006)}]{G2_star}
Sulaksono, A. and Mart, T. (2006).
\newblock Low density instability in relativistic mean field models.
\newblock \emph{Phys. Rev. C} 74, 045806.
\newblock \doi{10.1103/PhysRevC.74.045806}
\bibAnnoteFile{G2_star}

\bibitem[{{Tews} et~al.(2017){Tews}, {Lattimer}, {Ohnishi}, and
  {Kolomeitsev}}]{tews2017}
{Tews}, I., {Lattimer}, J.~M., {Ohnishi}, A., and {Kolomeitsev}, E.~E. (2017).
\newblock {Symmetry Parameter Constraints from a Lower Bound on Neutron-matter
  Energy}.
\newblock \emph{\apj} 848, 105.
\newblock \doi{10.3847/1538-4357/aa8db9}
\bibAnnoteFile{tews2017}

\bibitem[{Tsang et~al.(2024)Tsang, Tsang, Lynch, Kumar, and
  Horowitz}]{Tsang_2024}
Tsang, C.~Y., Tsang, M.~B., Lynch, W.~G., Kumar, R., and Horowitz, C.~J.
  (2024).
\newblock Determination of the equation of state from nuclear experiments and
  neutron star observations.
\newblock \emph{Nature Astronomy} 8, 328–336.
\newblock \doi{10.1038/s41550-023-02161-z}
\bibAnnoteFile{Tsang_2024}

\bibitem[{{Tsang} et~al.(2009){Tsang}, {Zhang}, {Danielewicz}, {Famiano}, {Li},
  {Lynch} et~al.}]{2009PhRvL.102l2701T}
{Tsang}, M.~B., {Zhang}, Y., {Danielewicz}, P., {Famiano}, M., {Li}, Z.,
  {Lynch}, W.~G., et~al. (2009).
\newblock {Constraints on the Density Dependence of the Symmetry Energy}.
\newblock \emph{\prl} 102, 122701.
\newblock \doi{10.1103/PhysRevLett.102.122701}
\bibAnnoteFile{2009PhRvL.102l2701T}

\bibitem[{{Typel} and {Brown}(2001)}]{skin2}
{Typel}, S. and {Brown}, B.~A. (2001).
\newblock {Neutron radii and the neutron equation of state in relativistic
  models}.
\newblock \emph{\prc} 64, 027302.
\newblock \doi{10.1103/PhysRevC.64.027302}
\bibAnnoteFile{skin2}

\bibitem[{{Typel} and {Wolter}(1999)}]{tw99}
{Typel}, S. and {Wolter}, H.~H. (1999).
\newblock {Relativistic mean field calculations with density-dependent
  meson-nucleon coupling}.
\newblock \emph{\nphysa} 656, 331--364.
\newblock \doi{10.1016/S0375-9474(99)00310-3}
\bibAnnoteFile{tw99}

\bibitem[{{Vautherin} and {Brink}(1972)}]{skyrmea}
{Vautherin}, D. and {Brink}, D.~M. (1972).
\newblock {Hartree-Fock Calculations with Skyrme's Interaction. I. Spherical
  Nuclei}.
\newblock \emph{\prc} 5, 626--647.
\newblock \doi{10.1103/PhysRevC.5.626}
\bibAnnoteFile{skyrmea}

\bibitem[{{Vida{\~n}a} et~al.(2009){Vida{\~n}a}, {Provid{\^e}ncia}, {Polls},
  and {Rios}}]{2009Isaac}
{Vida{\~n}a}, I., {Provid{\^e}ncia}, C., {Polls}, A., and {Rios}, A. (2009).
\newblock {Density dependence of the nuclear symmetry energy: A microscopic
  perspective}.
\newblock \emph{\prc} 80, 045806.
\newblock \doi{10.1103/PhysRevC.80.045806}
\bibAnnoteFile{2009Isaac}

\bibitem[{Wiringa and Pieper(2002)}]{Wiringa_2002}
Wiringa, R.~B. and Pieper, S.~C. (2002).
\newblock Evolution of nuclear spectra with nuclear forces.
\newblock \emph{Phys. Rev. Lett.} 89, 182501.
\newblock \doi{10.1103/PhysRevLett.89.182501}
\bibAnnoteFile{Wiringa_2002}

\bibitem[{Wiringa et~al.(1995)Wiringa, Stoks, and Schiavilla}]{v18}
Wiringa, R.~B., Stoks, V. G.~J., and Schiavilla, R. (1995).
\newblock {An accurate nucleon-nucleon potential with charge independence
  breaking}.
\newblock \emph{Phys. Rev. C} 51, 38--51.
\newblock \doi{10.1103/PhysRevC.51.38}
\bibAnnoteFile{v18}

\bibitem[{Xia et~al.(2022)Xia, Maruyama, Li, Yuan~Sun, Long, and
  Zhang}]{Xia_2022}
Xia, C.-J., Maruyama, T., Li, A., Yuan~Sun, B., Long, W.-H., and Zhang, Y.-X.
  (2022).
\newblock Unified neutron star {EOS}s and neutron star structures in {RMF}
  models.
\newblock \emph{Communications in Theoretical Physics} 74, 095303.
\newblock \doi{10.1088/1572-9494/ac71fd}
\bibAnnoteFile{Xia_2022}

\bibitem[{Yakovlev and Pethick(2004)}]{Yakovlev_2004}
Yakovlev, D. and Pethick, C. (2004).
\newblock Neutron star cooling.
\newblock \emph{Annual Review of Astronomy and Astrophysics} 42, 169–210.
\newblock \doi{10.1146/annurev.astro.42.053102.134013}
\bibAnnoteFile{Yakovlev_2004}

\bibitem[{{Zhang} et~al.(2017){Zhang}, {Cai}, {Li}, {Newton}, and
  {Xu}}]{2017Zhang}
{Zhang}, N.-B., {Cai}, B.-J., {Li}, B.-A., {Newton}, W.~G., and {Xu}, J.
  (2017).
\newblock {How tightly is nuclear symmetry energy constrained by unitary Fermi
  gas?}
\newblock \emph{Nucl. Sci. Tech.} 28, 181
\bibAnnoteFile{2017Zhang}

\bibitem[{{Zuo} et~al.(2002){Zuo}, {Lejeune}, {Lombardo}, and {Mathiot}}]{zuo1}
{Zuo}, W., {Lejeune}, A., {Lombardo}, U., and {Mathiot}, J.~F. (2002).
\newblock {Microscopic three-body force for asymmetric nuclear matter}.
\newblock \emph{Eur. Phys. J. A} 14, 469--475
\bibAnnoteFile{zuo1}

\end{thebibliography}

\end{document}